\documentclass[aps,prb,a4paper,reprint,superscriptaddress,showpacs]{revtex4-1}
\usepackage{graphicx,graphics,color,epsfig}
\usepackage{amsmath}
\usepackage{amsfonts}
\usepackage{amssymb}
\usepackage{appendix}

\begin{document}
\title{Giant Goos-H\"{a}nchen Shift in Graphene Double-barrier Structures}

\author{Yu Song}
\email{kwungyusung@gmail.com}
\affiliation{Department of Physics and State Key Laboratory of Low-Dimensional
Quantum Physics, Tsinghua University, Beijing 100084, People's Republic of China}

\author{Han-Chun Wu}
\affiliation{CRANN and School of Physics, Trinity College Dublin, Dublin 2, Ireland}

\author{Yong Guo}
\affiliation{Department of Physics and State Key Laboratory of Low-Dimensional
Quantum Physics, Tsinghua University, Beijing 100084, People's Republic of China}

\begin{abstract}
We report giant Goos-H\"{a}nchen shifts [F. Goos and H. H\"{a}nchen,
Ann. Phys. \textbf{436}, 333 (1947)] for electron beams
tunneling through graphene double barrier structures.
We find that inside the transmission gap for the single barrier,
the shift displays sharp
peaks with magnitudes up to the order of electron beam width and
rather small full-widths-at-half-maximum,
which may be utilized to design valley and spin beam splitters with
wide tunability and high energy resolution.
We attribute the giant shifts to quasibound states in the structures.
Moreover, an induced energy gap in the dispersion can increase
the tunability and resolution of the splitters.
\end{abstract}
\date{\today}
\maketitle

In optics, it is well-known that a light beam undergoes a lateral shift
when it is totally reflected from a dielectric interface.\cite{define}
This phenomena is referred to as the Goos-H\"{a}nchen (GH) shift
\cite{GH} and can be theoretically explained based on the reshaping
of the wave packet.\cite{stationary}
Analogies of the GH shift have
been widely considered in various fields, including acoustics,\cite{acoustic}
electronics,\cite{qm,e1,spin} relativistic corrections,\cite{e2}
atomic optics,\cite{atomoptics} and neutronics.\cite{neutron}

Recently, the analogy for the massless electron has become of interest\cite{zhao,gGH,grapheneSB,jpcm,valley,valleysplitter}
since discovery of graphene, a monolayer of $sp^2$ bonded carbon atoms.\cite{graphene_realize}
It has been reported that the GH shift plays an important role in
the group velocity of quasiparticles along interfaces of graphene $p$-$n$ junctions,\cite{zhao,gGH}
whereby a twofold degeneracy on top of the usual spin and valley
degeneracies can be introduced\cite{gGH} and
coherent buffers and memories can be realized in graphene $p$-$n$-$p$ waveguides.\cite{zhao}
The valley-dependent GH shift\cite{valley,valleysplitter} based on
strained graphene has been very recently utilized by Zhai \emph{et al.} to design
a valley beam splitter.\cite{valleysplitter}
We note that,
to effectively realize the proposed splitter, the
difference of GH shifts for valley $K$ and $K'$ should be larger than
the longitudinal width of the electron beam defined as $w_y=w_b\cos^{-1}\bar{\alpha}$, where
$w_b$ and $\bar{\alpha}$ are the waist width and incident angle of the
electron beam, respectively.
This condition also ensures the validity of the stationary-phase approximation,\cite{CFL}
which is widely used in GH shift calculation\cite{zhao,gGH,grapheneSB,jpcm,valley,valleysplitter}
especially for resonant transmission.\cite{pre}
Considering a typical beam divergence
($\delta\bar{\alpha}\equiv\lambda_F/\pi w_b$
with $\lambda_F$ the Fermi wave-length)
of $1^{\circ}$-$0.1^{\circ}$,\cite{pre}
$w_b$ is $180$$\lambda_F$-$1800$$\lambda_F$.
In a $n$-$p$-$n$ single barrier structure (SBS),\cite{valleysplitter} this
condition may not be met without properly selecting the structural parameters.
On the other hand, the spin-dependent GH shift was also proposed to spatially split spin beams,
based on a SBS formed by a local magnetic field\cite{spinsymmetry}
and an electrostatic potential in a two-dimensional electron gas (2DEG).\cite{spin}
However, the displacements of the spin-dependent GH shifts are also found to
be insufficient.\cite{spin}
Therefore, finding a suitable structure to effectively implement these
ideas is one of the main issues in such an exciting field.

In this letter, we report a giant GH shift of electron
beams tunneling through a double barrier structure
(DBS) in graphene. We find that inside the transmission
gap (TG) for the constituted single barrier\cite{TG} the GH shift displays sharp peaks which are
absent in the SBS cases and were attributed to the quasibound
states formed in the DBS. Remarkably, we find that the
magnitudes of the peaks are much larger than the maximum
magnitude in the corresponding SBS and can easily
achieve the order of $w_y$. Together
with the rather small full-widths at half-maximum (FWHM) of the peaks, this giant GH shift can
be used for designing valley and spin beam splitters with
wide tunability and high energy resolution. We also investigate the effects of
the structural asymmetry of the DBS and the induced energy
gap in the linear dispersion. The results show that
the former suppresses the GH shift while the latter
enhances it, which increases the controllability of
devices based on the GH shift.

\begin{figure}
\centering
\includegraphics[width=\linewidth]{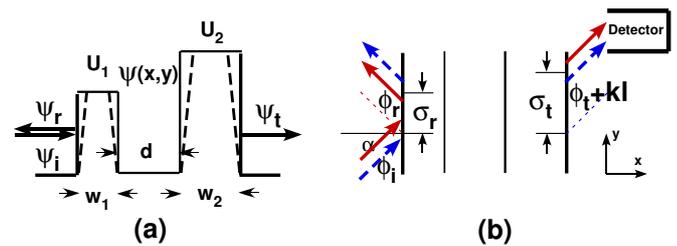}
\caption{(color online)
(a) Sectional and (b) top schematic diagrams for a particle quantum tunneling through a graphene DBS,
with the two barriers of width $w_{1(2)}$, height $U_{1(2)}$, and
distance $d$ between them. $l=w_1+w_2+d$ is the total width of the structure.
In (a), the dashed lines show smooth electric potentials with distributions
of error functions and transition regions' widths of $0.1 w_{1(2)}$.
In (b), the upper (red solid) and lower (blue dashed) components have their locus at
$\pm\delta/2$, $\sigma_r\mp\delta/2$, and $\sigma_t\pm\delta/2$
for the incident, reflected and transmitted beams, respectively.
A detector placed in a proper position of the outgoing region can detect
the giant GH shift.}
\end{figure}

To investigate the GH shift in graphene DBSs, we consider an incident
beam well collimated around some transverse wave vector $\bar{q}\in(-q_m,q_m)$ ($q_m=E/\hbar v_F$)
[corresponding to $\bar{\alpha}=\arcsin(\bar{q}/q_m)$],
quantum tunneling through a graphene DBS as sketched in Fig. 1.
The Dirac equation can be written as:
\begin{equation}
(v_{F}\boldsymbol{\sigma}\cdot \mathbf{p}+\Delta\sigma_{z}+U)\boldsymbol{\Psi}=E\boldsymbol{\Psi},
\end{equation}
where $v_F\approx10^6m/s$ is the Fermi velocity,
$\boldsymbol{\sigma}=(\sigma_x,\sigma_y)$ is the pseudospin operator
given by Pauli's matrices, $\mathbf{p}=(p_x,p_y)^T$ is the momentum operator,
$\Delta=mv_{F}^{2}(\Delta_{SO})$ is the energy gap owing to the sublattice
symmetry breaking\cite{gap1} (the spin-orbit interaction\cite{gap2}),
and $U=U_{1(2)}$ in the barriers while $U=0$ elsewhere.
The wave packets of the incident and associated reflected
and transmitted beams at the two terminals of the DBS can be given by
\begin{subequations}
\begin{equation}
\boldsymbol{\Psi}^{i}(x,y)=\frac{1}{\sqrt{2}}\int_{-q_m}^{q_m} dq f(q-\bar{q}) e^{ik(q)x+iqy}\left(
\begin{array}{cc}
e^{-i\alpha'(q)/2}\\
\lambda e^{i\alpha'(q)/2}
\end{array}\right),
\end{equation}
\begin{equation}
\boldsymbol{\Psi}^{r}=\frac{1}{\sqrt{2}}\int_{-q_m}^{q_m} dq r(q) f(q-\bar{q}) e^{-ik(q)x+iqy}\left(
\begin{array}{cc}
-ie^{i\alpha'(q)/2}\\
\lambda ie^{-i\alpha'(q)/2}
\end{array}\right),
\end{equation}
\begin{equation}
\boldsymbol{\Psi}^{t}=\frac{1}{\sqrt{2}}\int_{-q_m}^{q_m} dq t(q) f(q-\bar{q}) e^{ik(q)x+iqy}\left(
\begin{array}{cc}
e^{-i\alpha'(q)/2}\\
\lambda e^{i\alpha'(q)/2}
\end{array}\right).
\end{equation}
\end{subequations}
Here each plane wave of the spinor form is a solution of Eq. (1),
and a basis is used to ensure that the product of the upper and lower components
is real.\cite{gGH}
$f(q-\bar{q})$ is the angular spectral distribution
which can be assumed to be of Gaussian profile,
$w_ye^{-w_y^2(q-\bar{q})^2/2}$.
$k(q)=\sqrt{q_m^2-(\Delta/\hbar v_F)^2-q^2}$
is the longitudinal wave vector and $\alpha'(q)=\sin^{-1}(\hbar v_Fq/\sqrt{E^2-\Delta^2})$
is the phase angle.
Due to the induced energy gap $\Delta$, $\alpha'$ is usually different from the incident angle
and a factor $\lambda=\sqrt{E^2-\Delta^2}/(E+\Delta)$
appears in the lower component of the spinors.
For the reflected and transmitted beams,  $r(q) =|r(q)|e^{i\phi_r(q)}$ and
$t(q)=|t(q)|e^{i\phi_t(q)}$ are the reflection and transmission coefficients respectively,
which can be determined by the continuities of each component and calculated
by the standard transfer-matrix method.\cite{TMM}

\begin{figure}
\centering
\includegraphics[width=\linewidth]{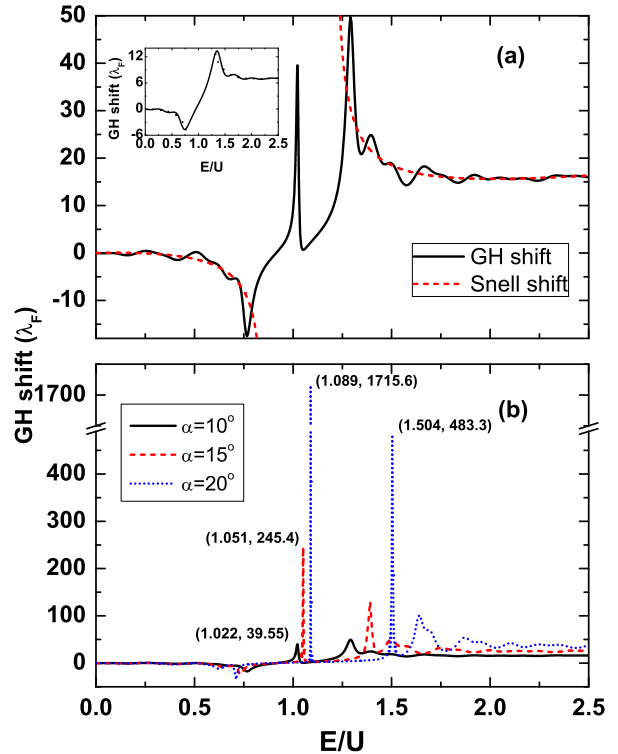}
\caption{(color online)
(a) The GH shift (solid) in a symmetric graphene DBS [$U=U_1=U_2=$62 meV, $w_1=w_2=$100 nm, $d=$50 nm]
as a function of the incident energy at $\bar{\alpha}=10^\circ$.
The insert shows the GH shift in a SBS with the same barrier height
and width.
The dashed line indicate the semi-classical shift predicted by the Snell's
law, which has no definition in the TG.
(b) The dependence of the shift sharp peak(s) on the incident angle,
$\bar{\alpha}=10^\circ,15^\circ,$ and $20^\circ$. The positions and heights of
the four giant GH shift peaks are marked.}
\end{figure}

From the stationary-phase approximation,\cite{CFL}
the vanishing of the gradient of the total phase in direction gives
the locus of the steady wave packet peak.
Due to the spinor nature of graphene, the loci of the two components
in a beam are found to be different.\cite{gGH}
For the incident and transmitted beam,
the upper component exceeds the lower one with a distance of $\delta$
($\delta=1/k$), while for the reflected beam the inverse becomes the case [see, Fig.1(b)].
The deviation between the loci of the reflected (transmitted) and the
incident beams thus gives corresponding GH shift,
$\sigma_r^\pm=-d\phi_r/dq|_{\bar{q}}\mp\delta$
and $\sigma_t^\pm=-d\bar{\phi}_t/dq|_{\bar{q}}$,
where $\bar{\phi}_t\equiv\phi_t+kl$. 
Note the GH shift in reflection is component dependent while
the shift in transmission is not [see, Fig.1(b)].
We'd like to use $\sigma_r=-(d\phi_r/dq)|_{\bar{q}}$ to describe the average shift in reflection.

Fig. 2(a) shows the calculated GH shift for a symmetric graphene DBS as a function
of $E$ at a fixed $\bar{\alpha}=10^{\circ}$.
In this calculation, $\Delta=0$.
To make sure the electron density of states coincides with a true system
and the system stays in a ballistic regime, typical values are used:
$U=62$ meV,
$w=100$ nm,
and
$d=$50 nm.
These parameters also ensure the legitimacy of the stationary-phase approximation.\cite{2Dgd,pre,CFL}
One can see from Fig. 2(a) that outside the TG,
the GH shift shows the same trend as the shift predicted by geometric
optics using Snell's law.\cite{grapheneSB,valleysplitter}
Moreover, the GH shift oscillates around the Snell shift as it is
enhanced (suppressed) at $k_iw_i=N\pi$ [$k_iw_i=(N-1/2)\pi$] with $N=1,2,3,...$ where
resonant (antiresonant) tunneling happens.
Note, the positive peak nearest the TG (denoting as $P_{SBS}$) which
has the maximum magnitude for tunneling through a SBS cannot
be enhanced to the order of the longitude beam width without properly selecting
the structural parameters.

Remarkably, a significantly sharp peak with a magnitude comparable to $P_{SBS}$
appears inside the TG [around $E/U=1$ in Fig. 2(a)].
This is absent in the case of electron beams tunneling through graphene SBSs [see the inset of Fig. 2(a)].
One can clearly see from Fig. 2(b) that with increasing incident angle
the peak value of the GH shift dramatically increases.
At $\bar{\alpha}=20^{\circ}$, the peak value reaches $\sim1700\lambda_F$
($\sim16000$ nm)
which is about ten times of the corresponding $P_{SBS}$
and is in the order of the longitudinal beam width.
Therefore, the obtained results suggest that
the valley splitter based on this structure can be realized with much
looser conditions, since giant GH shift for valley $K (K')$ can be obtained inside the
TG while the GH shift for valley $K' (K)$ retains the order of $\lambda_F$.
In addition, due to the rather small FWHM
the splitter will possess a much higher energy or wave vector resolution.

\begin{figure}
\centering
\includegraphics[width=\linewidth]{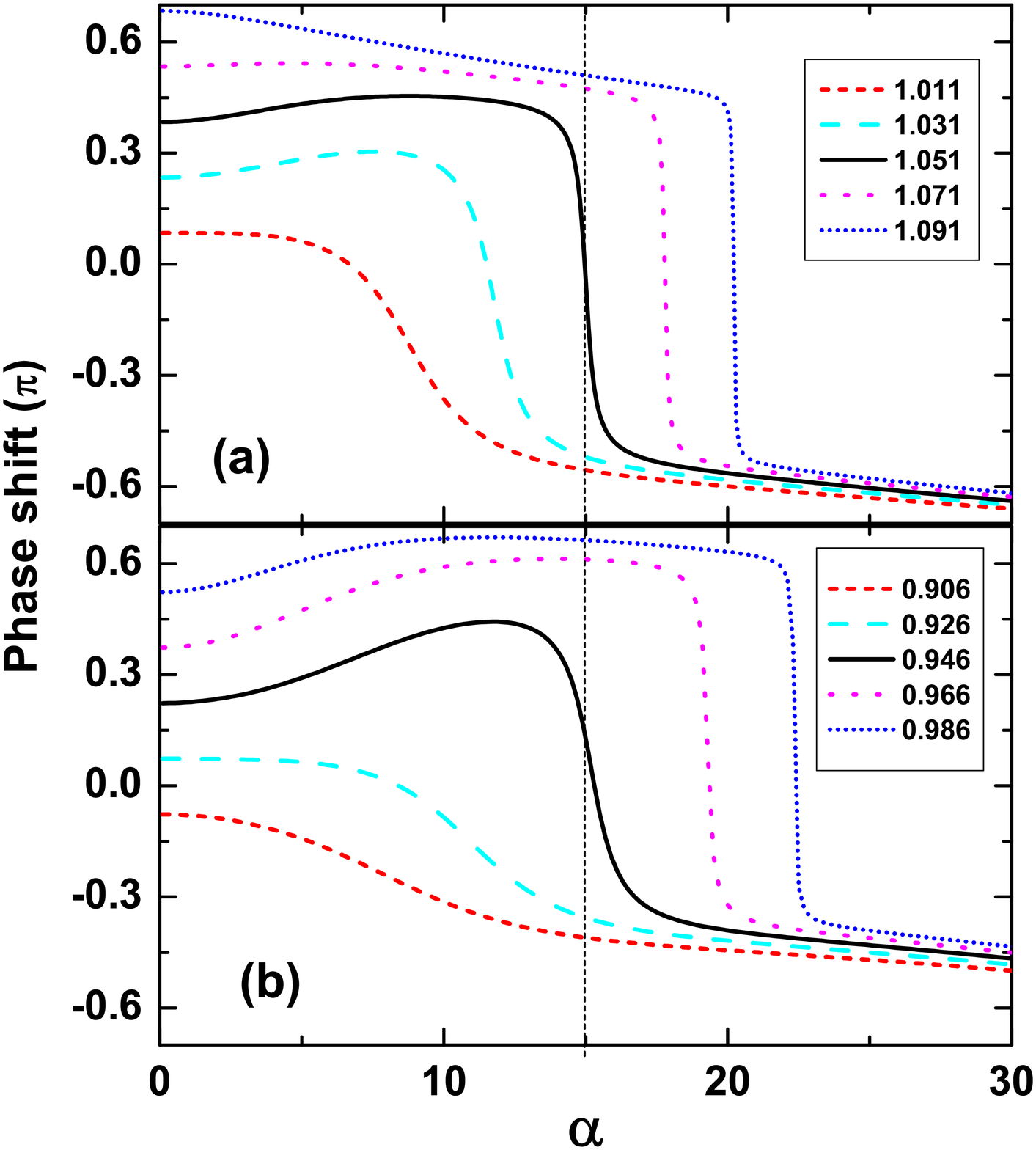}
\caption{(color online)
The phase shift of the transmitted beam vs. the incident
angle for beams with different $E/U$ shown in the figure.
(a)/(b) for a rectangular/smooth DBS.
}
\end{figure}

It is surprising that the GH shift in a graphene DBS possesses such
giant magnitudes,
since the shift along a single interface or through a SBS (inside the TG)
is of the order of Fermi wave-length $\lambda_F$.
\cite{gGH,zhao,grapheneSB,jpcm,valley,valleysplitter}
Here, we attribute it
to the quasibound states
in the DBS, which are formed by the evanescent waves in the two barriers.
It is well-known that 
plane waves with different $q$ generally present different phase shifts,
leading to the reshaping of the wave packet and thus the GH shift
of the electron beam.
In our case, when the center plane wave
is aligned to the quasibound state,
multiple interferences will arise through the quasistanding waves
between the two barriers.
This will lead to a remarkable
difference in the phase shifts between the center plane wave and adjacent ones
[see Fig. 3(a)].
Accordingly, a giant GH shift for such an electron beam is present
inside the TG.
One can also see clearly from Fig. 3(a) that with the increasing $E/U$ of the beam,
the incident angle which shows the biggest slope of $-\partial\phi_r/\partial\alpha$
also increases, which is consistent with Fig. 2(b).
Thus, the observed giant GH shift inside the TG is due to the
quasibound states formed in graphene DBSs.

\begin{figure}
\centering
\includegraphics[width=\linewidth]{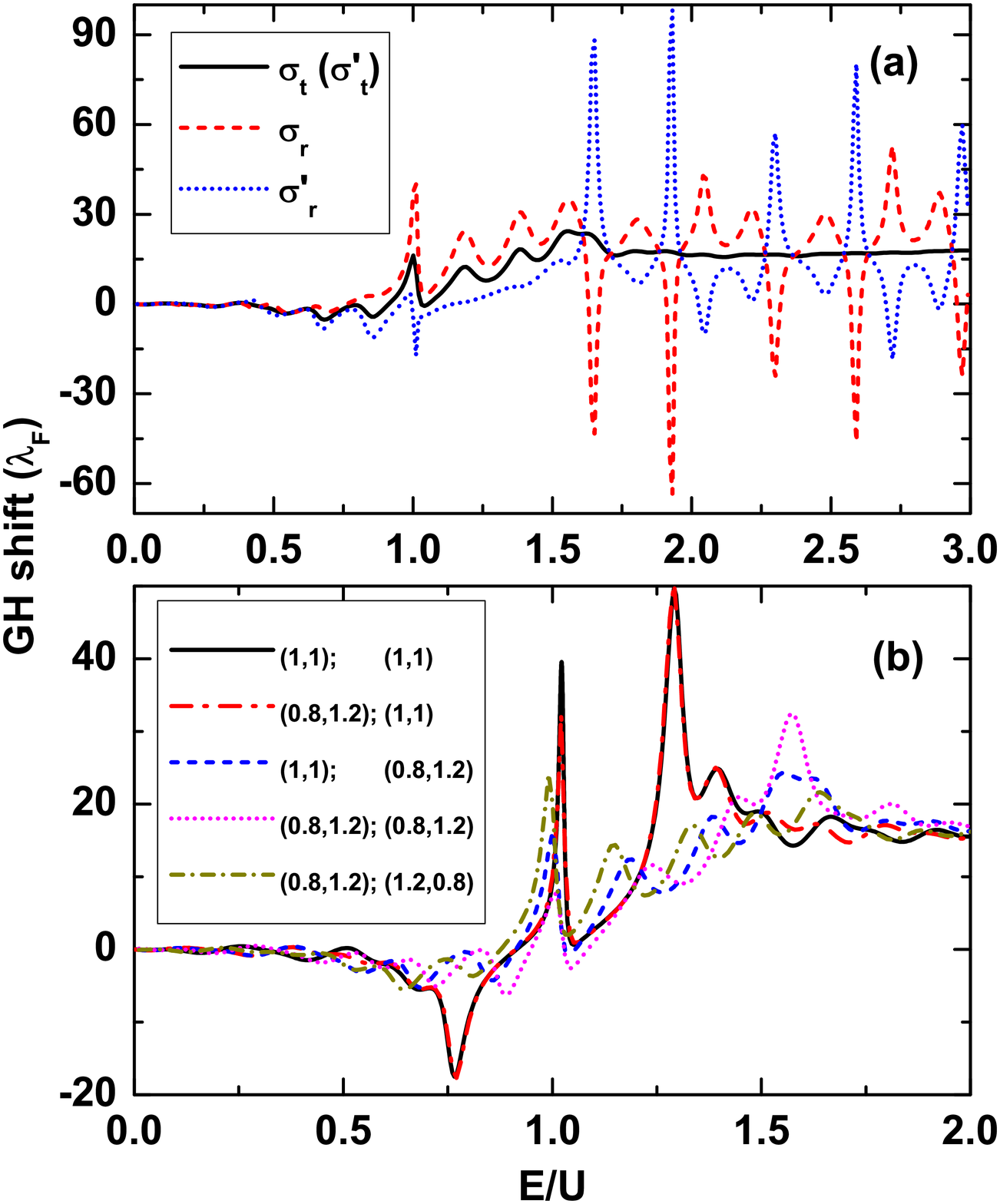}
\caption{(color online)
 (a) GH shifts in transmission and reflection for an asymmetric DBS with
$U=$62 meV, $U_1=0.8U$, $U_2=1.2U$, $w_1=w_2=w=$100 nm, and $d=$50 nm.
(b) GH shifts in transmission for different cases of structural asymmetries
with the parameters $(w_1/w,w_2/w)$ and $(U_1/U,U_2/U)$ in the sequence given in the figure.
For all cases, $\bar{\alpha}=10^{\circ}$.}
\end{figure}

One may wonder how the smoothness of realistic barriers will affect the
magnitude of the GH shift.
To answer this problem, we consider a realistic potential which varies smoothly on the scale of the graphene
lattice constant
and adopt a typical potential profile\cite{valleysplitter} of
$U(x)=0.5U[\textmd{erf}(2x/L_b-2)+\textmd{erf}(2(w-x)/L_b-2)]$,
where $\textmd{erf}(x)$ is the error
function and the width of the transition region is set as $L_b = 0.1w$ (see Fig. 1(a)).
We calculated the transmission coefficient, phase shift, and also the
GH shift. Similar results have been achieved. Fig. 3(b) shows the phase shifts as a function
of incident angle for a smooth DBS.
Compared with the case of a rectangular DBS with the same $\bar{\alpha}=15^\circ$, the giant GH shift
for a smooth DBS appears at a lower incident energy (from $E/U$=1.051 to 0.946) and the magnitude
of the giant GH peaks decreases from 245.4$\lambda_F$ to 67.4$\lambda_F$.
To achieve the GH shift in the order of the longitude beam width, one can increase the incident angle (about 23$^\circ$).
Thus the smoothness of the potential barriers will not restrict the
use of the giant GH shift.
For convenience, we will adopt the rectangular model in the following discussions.

\begin{figure}
\centering
\includegraphics[width=\linewidth]{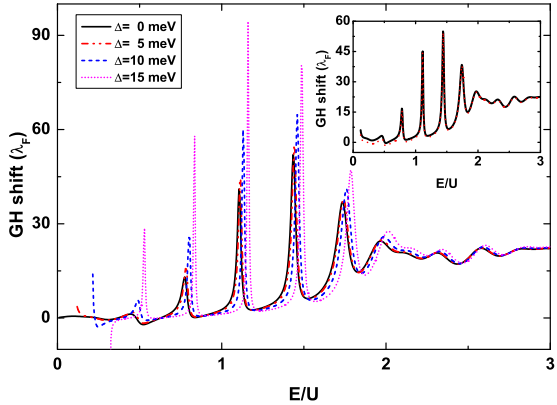}
\caption{(color online)
The GH shift in transmission through a graphene DBS with
$w_1=w_2=$50 nm, $d=$100 nm and $U=U_1=U_2=$50 meV.
The induced gap is $\Delta=0,5,10,15$ meV as indicated in the figure.
Insert: the GH shifts for the reflected and transmitted beams for $\Delta=5$ meV
in the same structure. For all cases, $\bar{\alpha}=20^{\circ}$.
}
\end{figure}

We now study the effect of the structural asymmetry of the DBS which may be
introduced during sample preparation and gate control,
i.e., the two barriers have different widths and/or heights.
Fig. 4 shows the GH shifts in transmission and reflection for
asymmetric DBSs.
Interestingly, one can see from Fig. 4(a) that the GH shift in reflection
crucially depends on which terminal the beam is being reflected from,
where $\sigma_r$ and $\sigma_t$ ($\sigma_r'$ and $\sigma_t'$)
mean the GH shift in reflection and transmission for electron beams
incident on the $U_1$ ($U_2$) terminal (as shown in Fig. 1).
We would like to point out that $\sigma_r$ and $\sigma_r'$ are
found to follow the relation of $\sigma_r+\sigma_r'=2\sigma_t$
which holds for any asymmetric DBSs and
can be understood by the relation between the
reflection and transmission coefficients.
Through the scattering matrix,\cite{TMM} we can get $t'=t$ and $r'=-tr^\ast/t^\ast$,
which imply 
$\bar{\phi}_t'=\bar{\phi}_t$ and $\phi_r+\phi_r'=2\bar{\phi}_t$, respectively.
$r$ ($r'$) and $t$ ($t'$) are the reflection and transmission coefficients
of the beam reflected from or tunneling through the $U_1$ ($U_2$) terminal,
where the term $kl$ has been contained in $t$ and $t'$.
Using the definition of GH shift, we get
$\sigma_t=\sigma_t'$ and $\sigma_r+\sigma_r'=2\sigma_t$.
Note for a symmetrical DBS, $\bar{\phi}_t=\phi_r$ ($\bar{\phi}_t'=\phi_r'$)
thus $\phi_r'=\phi_r$,
which indicates that the two GH shifts in reflection will become identical in this case.

We now take $\sigma_t$ ($\sigma_t'$) as a target to
evaluate the effect due to the structural asymmetries.
As shown in Fig. 4(b), the effect of structural asymmetry always
suppresses the GH shift inside the TG.
For the case of barriers with the same height but different widths,
the structural asymmetry only decreases the magnitude of
the GH shift inside the TG while not moving the peak position.
For the case of barriers of different heights, the effect of
structural asymmetry
not only decreases the magnitude of the GH shift inside the TG
but also makes the peak position shift.
Moreover, the FWHM of the peak of the GH shift increases.

Fig. 5 shows the GH shift for the transmitted beam in a graphene DBS
with different induced energy gaps in the linear dispersion.
As is seen, when there is a nonzero gap,
the GH shift has no definition for $E^2\cos^2\bar{\alpha}<\Delta^2$ as the
electron cannot propagate freely even in the non-modulated regions.
Meanwhile, the GH shifts for the reflected and transmitted beams
differ (see insert in Fig. 5).
With increasing  induced energy gap, the peak positions of the GH
shift inside the TG move to higher energies.
Moreover, the magnitudes of the peaks increase and the FWHMs become even smaller,
which increases the tunability and energy resolution of the valley or spin splitter device.
The underlying physics is that, the presence of energy gap increases the
modulus of the longitudinal wave vector ($\kappa$), which makes
a stronger multiple interference effect and thus a bigger GH shift at a little higher energy.

In summary, we have theoretically calculated the GH
shifts of reflected and transmitted electron beams in a graphene
DBS. Interestingly, we found that the
GH shift displays sharp peaks inside the TG for the constituted SB, which are
absent in the graphene SBS cases and can be attributed
to the quasibound states formed in the DBS.
The reported giant GH shift can be detected
in the transmitted beam by placing a detector in the
outgoing region and far away from the incident position (see Fig. 1(b)).
In the TG, the detector will collect no electrons unless
the incident energy is aligned to the quasibound states.
The results obtained in this work suggest the feasibility of making
valley splitter based on graphene DBSs.
The tunability and energy resolution of the splitter
can be further increased by an induced energy gap in the linear dispersion.
By the spin-dependent giant GH shift,
a spin beam splitter restricted by the small magnitude of the GH
shift\cite{spin} may be realized in 2DEG or graphene based DBSs now.

This work was supported by the NSFC (10974109 and 11174168),
the SRFDP (20100002110079), and the 973 Program (2011CB606405).
HCW was grateful to the SFI Short Term Travel fellowship support
during his stay at PKU and thanks Dr. Cormac \'{O} Coile\'{a}in
for reading the paper and giving valuable comments.

\end{document}